\documentclass[iop]{emulateapj}

\newcommand{\HII}{{\ion{H}{2}}}

\newcommand{\OIIIHb}{[{\ion{O}{3}}]/H$\beta$}
\def\ratioR23{([\ion{O}{2}]~$\lambda$3727 +[\ion{O}{3}]~$\lambda\lambda$4959,5007)/H$\beta$}
\def\R23{${\rm R}_{23}$}

\newcommand{\NII}{[{\ion{N}{2}}]}

\newcommand{\OIIIOII}{[\ion{O}{3}]/[\ion{O}{2}]}

\newcommand{\OH}{$\log({\rm O/H})+12$}

\newcommand{\NIIHa}{[\ion{N}{2}]/H$\alpha$}
\newcommand{\SIIHa}{[\ion{S}{2}]/H$\alpha$}

\newcommand{\SII}{[{\ion{S}{2}}]}

\newcommand{\Hb}{{H$\beta$}}
\def\O4363{[{\ion{O}{3}}]~$\lambda$4363}
\newcommand{\OIII}{[{\ion{O}{3}}]}

\def\L60{L$_{60}$}

\newcommand{\logM}{$\log({\rm M}/{\rm M}_{\odot})$}

\shorttitle{}
\shortauthors{}

\begin{document}

\title{A rise in the ionizing photons in star-forming galaxies over the past 5 billion years}

\author{Lisa J. Kewley\altaffilmark{1,2}, H. Jabran Zahid\altaffilmark{3}, Margaret J. Geller\altaffilmark{3}, Michael A. Dopita\altaffilmark{1,5}, Ho Seong Hwang\altaffilmark{4}, \& Dan Fabricant\altaffilmark{3}}
\email {kewley@ifa.hawaii.edu}
\altaffiltext{1}{RSAA, Australian National University, Cotter Road, Weston Creek, ACT 2611, Australia}
\altaffiltext{2}{Institute for Astronomy, University of Hawaii, 2680 Woodlawn Drive, Honolulu, HI, USA}
\altaffiltext{3}{Smithsonian Astrophysical Observatory, Cambridge, MA 02138, USA}
\altaffiltext{4}{School of Physics, Korea Institute for Advanced Study, 85 Hoegiro, Dongdaemun-Gu, 130-722 Seoul, Korea}
\altaffiltext{5}{Astronomy Department, King Abdulaziz University, P.O. Box 80203, Jeddah, Saudi Arabia}

\begin{abstract}

We investigate the change in ionizing photons in galaxies between $0.2<z<0.6$ using the F2 field of the SHELS complete galaxy redshift survey.  We show, for the first time, that while the \OIIIHb\ and \OIIIOII\ ratios rise, the \NIIHa\ and \SIIHa\ ratios fall significantly over the $0.2<z<0.35$ redshift range for stellar masses between $9.2< \log ({\rm M / M}_{\odot}) <10.6$.  The \OIIIHb\ and \OIIIOII\ ratios continue to rise across the full $0.2<z<0.6$ redshift range for stellar masses between $9.8< \log ({\rm M / M}_{\odot}) <10.0$.  We conclusively rule out AGN contamination, a changing ISM pressure, and a change in the hardness of the EUV radiation field as the cause of the change in the line ratios between $0.2<z<0.35$.  We find that the ionization parameter rises significantly with redshift (by 0.1 to 0.25 dex depending on the stellar mass of the sample).  We show that the ionization parameter is strongly correlated with the fraction of young-to-old stars, as traced by the \Hb\ equivalent width.  We discuss the implications of this result on higher redshift studies, and we consider the implications on the use of standard optical metallicity diagnostics at high redshift.
\end{abstract}


\section{Introduction}
An understanding of the conditions under which stars formed at all epochs is critical in constraining theoretical models of galaxy evolution. 
Recent observations of $z>1$ star-forming galaxies suggest that the ISM state and/or the hardness of the EUV radiation field was more extreme in the past than in the present day.  Galaxies at high redshift display larger line ratios (\OIIIHb\ and/or \NIIHa) than local galaxies \citep[e.g.,][and references therein]{Hainline09,Bian10,Yabe12,Kewley13b,Holden14,Steidel14}.   These enlarged line ratios have been interpreted in terms of a contribution from an AGN or shocks \citep{Groves06,Wright10,Trump11,Forster13,Newman14}, a larger nitrogen abundance \citep{Masters14}, a larger ionization parameter \citep{Brinchmann08b,Steidel14}, and/or a higher ISM pressure \citep{Shirazi13}.  Measurements of the ionization parameter and electron density of the ISM using rest-frame optical line ratios support the general picture of a larger ionization parameter or electron density -- although the useable samples are small \citep{Liu08,Hainline09,Bian10,Nakajima12,Shirazi13,Shirazi13b,Nakajima14}.  However, some high redshift galaxies have electron densities similar to local galaxies \citep[see][]{Rigby11,Bayliss13}, implying a different cause of the anomalous line ratios.

\citet{Kewley13a} used theoretical simulations to show how the \NIIHa\ and \OIIIHb\ ratios of star forming galaxies and AGN may change with redshift, given four sets of extreme assumptions.  If the ISM conditions change with redshift, then the standard optical classification schemes and other line ratio diagnostics may not be applicable at high redshift \citep{Kewley13b}.  Here, we investigate the change in optical line ratios across the intermediate redshift range $0.2<z<0.6$ using the F2 field of the SHELS galaxy redshift survey.  The SHELS survey is 95\% complete to a limiting magnitude of ${\rm R} = 20.6$, allowing the optical line ratios to be investigated as a function of redshift and stellar mass.   At high redshift, emission-line selection effects are difficult to remove.  Galaxies with low \OIII\ or \NII\ luminosities may be missing from high-redshift samples, potentially accounting for a large fraction of the observed change in line ratios with redshift \citep{Juneau14}.   The high level of spectroscopic completeness of the SHELS sample avoids these issues and allows one to robustly constrain the change in line ratios with time, as well as the cause(s) of the change in line ratios.  In this paper, we use the SHELS sample to show that the \OIIIHb, \NIIHa, and \SIIHa\ ratios change systematically from $z=0.2$ to $z=0.6$, and we demonstrate that this change is caused by a change in the number of ionizing photons per unit area within galaxies, as a function of redshift.   Throughout this paper, we we use the flat $\Lambda$-dominated cosmology as measured by the 7 year WMAP experiment \citep[$h=0.7$, $\Omega_{m}=0.3$;][]{Komatsu11}.

\section{Sample and Derived Quantities}\label{Sample}

The SHELS survey is based on two of the Deep Lens Survey Fields \citep{Wittman06}.  We use the F2 field from the SHELS survey, 
described in \citet{Geller14}.  The 4 square degree field contains 12,705 spectroscopic redshifts for galaxies with r$\leq$20.6. The survey is 95\% complete to this limit.

The SHELS spectra were obtained by \citet{Geller05} using the Hectospec multi-fiber spectrograph on the MMT \citep{Fabricant98,Fabricant05}.   The spectra cover the full optical wavelength range from 3700 - 9100 \AA, at a spectral resolution of $\sim 5$\AA.  This high spectral resolution is sufficient to resolve the \SII\ doublet, allowing electron density estimates to be made by stacking in bins of stellar mass and redshift.  Example individual spectra are shown in \citet{Geller14}.

We have selected a sub-sample of the SHELS catalog according to the following criteria:
\begin{itemize}
\item A redshift in the range $0.2<z<0.6$.  The lower redshift limit avoids aperture effects, which can be large for $z\lesssim 0.2$ \citep{Kewley04}.
\item The 4000\AA\ break index, ${\rm D}_{n} 4000 < 1.5$.  This selection limits the sample to stellar ages $\lesssim 1$~Gyr, and is insensitive to the metallicity of the stellar population \citep[e.g.,][]{Poggianti97}.  
\end{itemize}

The ${\rm D}_{n} 4000$ index is defined as the ratio of the flux ($f_{\nu}$) in the $4000 - 4100$\AA\ and  $3850 -  3950$\AA\ bands \citep{Balogh99}.  The ${\rm D}_{n} 4000$ index has been shown to be less sensitive to reddening effects and contains less uncertainties from the stellar continuum than the original ${\rm D} 4000$ index \citep{Balogh99}.    The 4000\AA\ break is produced mainly by the heavy element absorption in the atmospheres of the older stellar population- the hotter and younger stars produce a smooth continuum in this region of the spectrum. Thus, the ${\rm D}_n 4000$ index provides a relative measure of the age of the stellar population.  The ${\rm D}_{n} 4000$ distribution for SHELS is described in \citet{Fabricant08} and \citet{Geller14}.  \citet{Geller14} showed that the SHELS ${\rm D}_{n} 4000$ distribution is bimodal, with ${\rm D}_{n} 4000<1.5$ selecting for actively star-forming galaxies and ${\rm D}_{n} 4000>1.5$ selecting for quiescent galaxies.

\subsection{Stacking Method}

To avoid emission-line selection effects, we stack the data in bins of stellar mass, as described in \citet{Geller14}.  We divide our sample into bins of stellar mass in $0.2$~dex increments. At least $\sim 50$ galaxies per bin are required to provide measurable \OIII\ and \Hb\ emission-lines in each stacked spectrum. The spectra and observational uncertainties are linearly interpolated to a common rest-frame wavelength vector based on the redshift of the bin.  The interpolated rest-frame wavelength vector has a spectral resolution of 1.5\AA\ and spans an observed wavelength range of $\lambda3500-9100$~\AA.   The average flux at each resolution element is taken and the errors are added in quadrature.  \citet{Geller14} conducted an in-depth analysis of the use of stacking for the SHELS sample.  They showed that stacking does not bias line ratio analysis for SHELS.

\subsection{Stellar Masses}\label{Mass}

Stellar masses were derived for our sample from the broad-band photometry using the {\it  Le Phare} code by Arnouts \& Ilbert (private communication), with the \citet{Bruzual03} stellar templates as inputs.  These stellar templates were given for two metallicities and for seven exponentially decreasing star formation models (${\rm SFR} \propto e^{t/ T}$) with $T = 0.1, 0.3, 1, 2, 3, 5, 10, 15, 30$~Gyr over a stellar population age between $0-13$~Gyr.   To correct for dust attenuation, we apply the \citet{Calzetti00} law with E(B-V)$=0 - 0.6$.  The {\it  Le Phare} code provides a mass probability distribution.  We use the median of the mass distribution as the stellar mass of our galaxies.

\subsection{Ionization Parameter}

The ionization parameter, $q$, is defined as the number of hydrogen ionizing photons passing through a unit area per second divided by the number density of hydrogen atoms.  The volume-weighted mean ionization parameter, $q$, can be defined in terms of the Str\"{o}mgren radius $R_{s}$ for a filled spherical \HII\ region as 

\begin{equation}
q = \frac{2^{2/3}S_{*}(t)}{4 \pi R_{s}^2 n}
\end{equation}

where $n$ is the total number density, and $S_{*}(t)$ is the number of ionizing photons produced by the exciting stars. 

A larger ionization parameter at high redshift has been proposed as a potential cause of the change in line ratios with redshift \citep[e.g.,][]{Brinchmann08b,Shirazi13,Steidel14}.  However, most high redshift galaxy spectra do not contain sufficient emission-lines to reliably constrain the ionization parameter.  Fortunately, we are able to constrain the ionization parameter for SHELS, and to investigate how the ionization parameter changes both as a function of redshift and as a function of stellar mass.

We measure the ionization parameter using the \OIIIOII\ ratio using the theoretical iterative approach described in \citet{Kewley02}, and updated in \citet{Kobulnicky04}.   We use the equivalent width ratio of \OIIIOII\ in these calculations.  \citet{Kobulnicky03a} and \citet{Zahid11} showed that the \OIIIOII\ flux ratios are well represented by the equivalent width ratio of \OIIIOII .   For redshifts $z<0.4$, we have verified that the use of equivalent widths yields the same results as using extinction-corrected line fluxes.  

\section{AGN contamination}\label{AGN}

An AGN produces a hard ionizing radiation field which leads to elevated \NIIHa\ and \OIIIHb\ line ratios.  A larger relative contribution from an AGN at high redshift has been proposed as the cause of the rise in \OIIIHb\ ratio with redshift \citep{Groves06,Wright10,Trump11}.   Sample selection is likely to play a major role in whether AGN contribute to the observed line ratios in a given sample.  

\begin{figure}[!t]
\epsscale{1.2}

\plotone{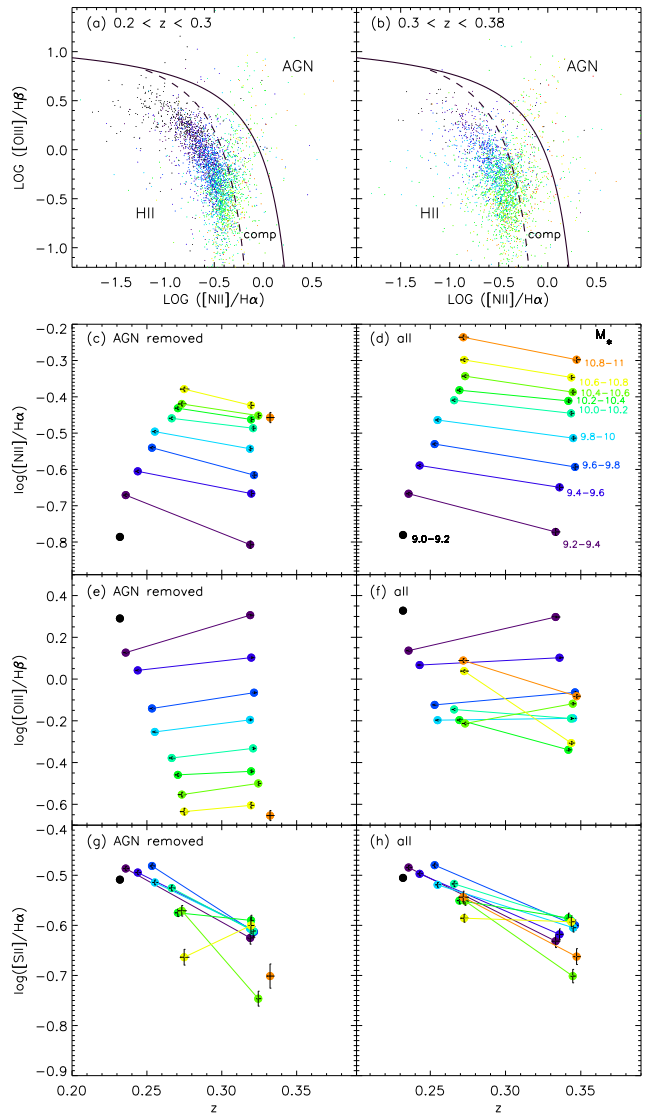}
\caption[AGN_removal]{(Upper panel) the optical diagnostic BPT diagram showing how AGN are removed from the SHELS $0.2<z<0.38$ sample.  Both AGN and composites are removed, yielding a purely star-forming sample.  The middle panels show the the \NIIHa, and \OIIIHb line ratios versus redshift for the SHELS stacks between $0.2<z<0.38$ where AGN have been removed (left panel) and AGN are included in the stacks (right panel).  The lower panels give the \SIIHa\ ratio versus redshift for the SHELS stacks between $0.2<z<0.35$ where AGN have been removed (left panel) and AGN are included in the stacks (right panel).  The colored curves correspond to the stacked spectra in stellar mass ranges, as shown in panel d.  The errors in the line ratios and the standard error of the redshift for each stack (shown) are smaller than the filled circles, thanks to the large number of galaxies in each stacked spectrum ($>500$).  AGN strongly affect the \OIIIHb\ ratios of the larger stellar mass stacks (\logM$>10.0$).  
\label{AGN_removal}}
\end{figure}

The effect of residual AGN contamination on stacked data is unknown.  However, the SHELS sample also allows us to investigate the effect of AGN on the \NIIHa\ and \OIIIHb\ optical line ratios ratios for galaxies between $0.2<z<0.38$ and on the \SIIHa\ ratio for galaxies between $0.2<z<0.35$.  We remove AGN and composites using the \OIIIHb\ versus \NIIHa\ optical diagnostic diagram, forming a purely star-forming sample \citep{Kewley06a} (Figure~\ref{AGN_removal} top panel)).  

In Figure~\ref{AGN_removal} (middle and lower panels), we show how the optical line ratios change with redshift for each stellar mass bin (colored curves).  The right panel shows the effect of AGN contamination on the line ratios in the stacked data.  The hard ionizing radiation field from an AGN strongly affects the \OIIIHb\ ratio because more ${\rm O}^{+}$ ions are ionized into ${\rm O}^{++}$ ions.  It is clear that if AGN are not removed, the \OIIIHb\ ratio is strongly affected by AGN for stellar masses larger than $\log {\rm M / M}_{\odot} > 10$.  Galaxies with lower stellar masses contain few AGN \citep[e.g.,][]{Groves06}.  Therefore, for $z \geq 0.4$ -- where we cannot remove the AGN -- we have restricted our analysis to galaxies with $\log ({\rm M / M}_{\odot}) < 10$ to minimise AGN contamination.

\section{The change in line ratios with redshift}\label{Ionization}

\begin{figure}[!t]
\epsscale{1.2}
\plotone{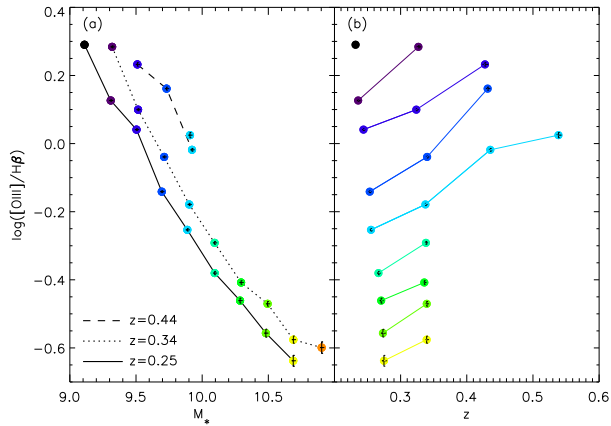}
\caption[OIIIHb_mass]{(a) The \OIIIHb\ ratio versus stellar mass for lines of constant redshift, as indicated in the legend.  (b) the \OIIIHb\ ratio versus redshift for lines of constant stellar mass.  The relationship between colors and stellar mass is shown in panel (a).  The \OIIIHb\ ratio rises by $0.2-0.3$ dex between redshifts $0.2<z<0.6$.  The change in \OIIIHb\ does not depend on stellar mass.
\label{OIIIHb_mass}}
\end{figure}

Figure~\ref{OIIIHb_mass} shows that the \OIIIHb\ ratio rises significantly between $0.2<z<0.6$. Although this ratio is also a strong function of stellar mass (Figure~\ref{OIIIHb_mass}a) nonetheless, the stellar mass does not appear to affect the size of the change in \OIIIHb\ shown in Figure~\ref{OIIIHb_mass}b for each mass bin.   The \OIIIHb\ ratio ratio rises by $0.28\pm0.01$~dex across the $0.2<z<0.6$ redshift range for all stellar masses in the range $9.8 < \log ({\rm M / M}_{\odot}) <10.0$.  This rise is consistent with high redshift studies which find large \OIIIHb\ line ratios in star-forming galaxies with $z>1.5$ \citep[][]{Hainline09,Bian10,Rigby11,Yabe12,Kewley13b,Holden14,Steidel14,Sanders14}. We will now briefly investigate possible mechanisms which could produce this observed  rise in the \OIIIHb\ ratio.

\subsection{Mean pressure in the \HII\ regions}\label{pressure}

Larger pressures in the \HII\ regions of galaxies may cause a rise in the \OIIIHb\ ratio.  This pressure can be inferred from the electron density of the gas.   In a fully ionized plasma with an isobaric density distribution, the particle density is defined in terms of the ratio of the mean ISM pressure,  and mean electron temperature, $T_e$, through $n =\frac{ P }{ T_e k }$, and $n$ is related to the electron density through $n=2 n_{e} (1+H_{e}/H)$.    For an ionized gas, the electron temperature is $\sim 10^4$~K and the electron density is therefore proportional to the ISM pressure \citep[see e.g.,][for a discussion]{Dopita06}.  We can trace the electron density in the SHELS sample for $z<0.35$ using the \SII~$\lambda 6717 /$\SII~$\lambda 6731$ ratio.  We find that the difference in \SII\ ratio between $0.2<z<0.35$ is consistent with no change with redshift; $\Delta($\SII~$\lambda 6717 /$\SII~$\lambda 6731$$ = 0.05 \pm 0.05$ from $z=0.2$ to $z=0.35$.  

Furthermore, a larger ISM pressure increases all three \NII, \SII, and \OIII\ ratios simultaneously because the larger pressure suppresses IR fine structure cooling, leading to more collisional excitations of all three \NII, \SII, and \OIII\ lines \citep[see Figure 2 in][]{Kewley13a}.  However, in the SHELS sample between $0.2<z<0.35$, the log(\NIIHa) and log(\SIIHa) ratios actually decrease with redshift by $\sim0.03 - 0.3$~dex (panels c and g of Figure~\ref{AGN_removal}).   We therefore rule out increasing pressure with redshift as the cause of the observed change in SHELS line ratios across the $0.2<z<0.35$ range.

\subsection{Hardness of the radiation field}\label{hardness}

Recently, a hard ionizing radiation field has been proposed as the cause of the large \OIIIHb\ line ratios in high redshift galaxies \citep{Steidel14,Stanway14}.  The effect of a hard ionizing radiation field on the optical line ratios is clearly illustrated in Figure~\ref{AGN_removal} by comparing the panels with and without an AGN for high mass galaxies (\logM$>10$).  A hard ionizing radiation field from either an AGN, shocks, or from a hot stellar population (such as Wolf-Rayet stars) ionizes more neutral nitrogen and sulphur into ${\rm N}^{+}$ and ${\rm S}^{+}$, as well as ionizing ${\rm O}^{+}$ into ${\rm O}^{++}$ \citep[e.g.,][]{Kewley01a,Levesque10}.  The \SII\ emission-line is particularly sensitive to the hardness of the ionizing radiation field.  In a hard ionizing radiation field, a large, warm and partially ionized zone containing ${\rm S}^{+}$ ion extends to the edge of the photoionised nebula, producing enhanced  \SIIHa\ ratios.  Because both \NIIHa\ and \SIIHa\ ratios fall with redshift, we can rule out a hard ionizing radiation as the cause of the change in SHELS line ratios with redshift.

\subsection{A changing ionization parameter with redshift}

\begin{figure}[!t]
\epsscale{1.2}
\plotone{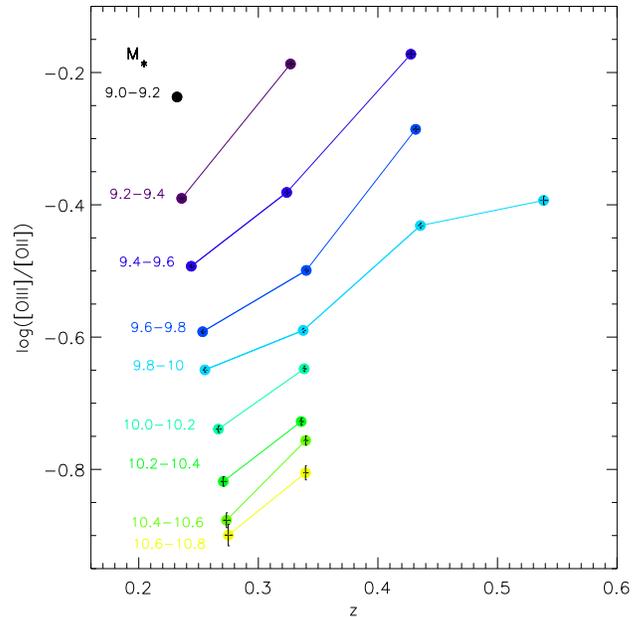}
\caption[OIIIOII_z]{The ionization-parameter sensitive \OIIIOII\ ratio as a function of redshift for our stacked data in bins of stellar mass (colored curves).  The \OIIIOII\ ratio rises with redshift for all stellar masses.  \label{OIIIOII_z}}
\end{figure}

A larger ionization parameter might produce the observed change in line ratios.  In Figure~\ref{OIIIOII_z}, we show that the ionization-parameter sensitive line ratio \OIIIOII\ rises with redshift (by up to $0.25$~dex).  This change in \OIIIOII\ ratio corresponds to a change in the ionization parameter of up to $\Delta \log(q) \sim 0.25$~${\rm cm~s}^{-1}$.  Although an increase in pressure can lead to an apparent rise in $q$ of this magnitude \citep{Dopita14}, the pressures needed would be in excess of $\log P/k > 6.0~{\rm K~cm}^{-3}$, which is excluded by an absence of any change in the  electron density as measured by the \SII~$\lambda 6717 /$\SII~$\lambda 6731$ ratio with redshift in the SHELS sample.  

Thus, any change in the hydrogen ionizing photon flux must be produced either by increasing the hardness (i.e. changing the slope) of the ionizing EUV radiation field, or by scaling up the whole ionizing radiation field by a some factor.  We have already ruled out a change in the hardness of the ionizing radiation field.  We now investigate the prospect of scaling the radiation field.

\begin{figure}[!t]
\epsscale{1.2}
\plotone{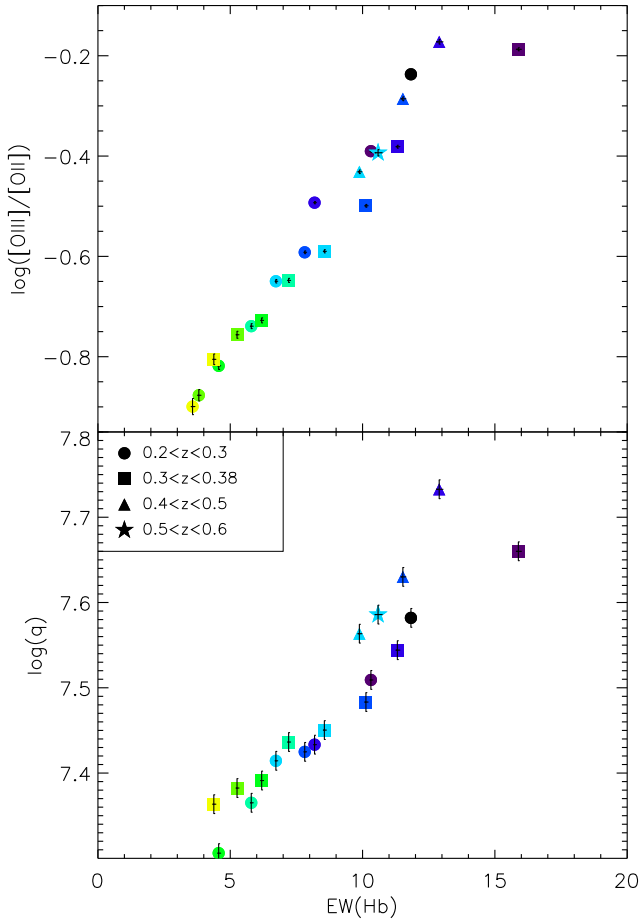}
\caption[logq_EWHb]{The \OIIIOII|\ ratio (upper panel) and the ionization parameter (log q; lower panel) versus the \Hb\ equivalent width for our SHELS sample in bins of stellar mass (colors as in Figure~1) and in bins of redshift (symbols) \label{logq_EWHb}}
\end{figure}

A larger hydrogen ionizing photon flux on the inner edge of the nebula could either be produced by collective effects such as a larger number of stars within an individual \HII\ region, or by geometrical effects such as a generally closer proximity of the stars to the ionisation front -- as in blister \HII\ regions.   If the first hypothesis is correct, we would expect  the ionization parameter to be related to the current star formation rate. This hypothesis can be tested by comparing the \OIIIOII\ ratio or the derived ionization parameter with the \Hb\ emission-line equivalent width, because the Balmer equivalent width traces the relative fraction of young stars to old stars in a galaxy \citep[see e.g.,][for a review]{Leitherer05}.  Indeed, we find a strong correlation (Figure~\ref{logq_EWHb}) between these quantities.  The Spearman-rank correlation coefficient between the ionization parameter and the \Hb\ equivalent width is 0.81, and the probability of obtaining this value by chance is formally zero ($6.6\times 10^{-5}$).  We conclude that a larger number of young to old stars within \HII\ regions is likely to be principally responsible for the change in ionization parameter with redshift in the SHELS sample.  

If the ionization parameter increases with redshift, then this will have important implications on the use of line ratios to derive metallicity at high redshift.  In particular, the \NIIHa\ ratio and the \OIIIHb\ ratios will be impacted strongly by a change in ionization parameter.  Both of these line ratios are in widespread use in high-z metallicity studies due to their insensitivity to extinction and flux calibration.   If the ionization parameter is not taken into account, the \NIIHa\ ratio will underestimate the metallicity.  The effect of ionization parameter on the \OIIIHb\ ratio is more complicated because the \OIIIHb\ ratio is double-valued with metallicity.  At high metallicity (\OH$>8.4$), the \OIIIHb\ metallicity will be underestimated, while at low metallicity (\OH$<8.4$), the \OIIIHb\ metallicity will be overestimated.  We recommend that the ionization parameter and metallicity be derived simultaneously to ensure reliable metallicity estimates for non-local galaxies.

\section{Conclusion}

We investigate the change in optical line ratios across $0.2<z<0.6$ for the SHELS survey.  For $9.8< \log ({\rm M / M}_{\odot}) <10.0$, the \OIIIHb\ and \OIIIOII\ ratios rise (by $\sim 0.2$~dex) across the full $0.2<z<0.6$ range.  For $0.2<z<0.35$ we show that while the \OIIIHb\ and \OIIIOII\ ratios rise, the \SIIHa\ and \NIIHa\ ratios fall significantly (by $0.05 -0.3$~dex) for stellar masses betweeon $9.2< \log ({\rm M / M}_{\odot}) <10.6$.  We explore these changes in terms of the electron density (or ISM pressure) of the gas, the slope of the ionizing radiation field, and the relative fraction of young to old stars.  We rule out a change in the electron density and a change in the hardness of the ionizing radiation field as dominant causes of our observed change in line ratios with redshift.

We examine the effect of AGN contamination.  We show that AGN substantially contaminates the \OIIIHb\ line ratios at stellar masses ($\log ({\rm M / M}_{\odot}) > 10$).  AGN removal is critical for high-z studies that interpret the \OIIIHb\ ratio in terms of star-forming galaxy properties.  However, contribution by AGN does not affect the general trends observed in the SHELS \NIIHa\ and \SIIHa\ line ratios.  It is unclear whether this lack of AGN sensitivity will hold at high-z, given potentially different ISM conditions and AGN properties.   

We show that the observed change in SHELS line ratios across $0.2<z<0.35$ is dominated by a rise in the ionization parameter with redshift.  The geometry, stellar mass contained, and the age of the stellar populations within \HII\ regions may all contribute to the observed ionization parameter.   The observed change in ionization parameter is strongly correlated with the fraction of young to old stars, as traced by the \Hb\ equivalent width.  

We emphasize that while the ionization parameter dominates the change in line ratios observed in SHELS, the ionization parameter may not be the dominant cause of a change in line ratios for different redshift ranges or different stellar mass ranges, particularly where shocks or AGN contamination may be present.

\acknowledgments

M.D. and L.K. are supported by an ARC Discovery Project DP130103925.  L.K. gratefully acknowledges an ARC Future Fellowship and the 2014 ANU Academic Women's Writing Workshop.  The Smithsonian Institution supports the research of M.G. and D.F.

\end{document}